\documentclass[twocolumn,pra]{revtex4}
\usepackage{amssymb}
\usepackage{graphicx}
\usepackage{dcolumn}
\usepackage{bm}
\usepackage{amsmath}

\newcommand{\bra}[1]{\langle #1 |}
\newcommand{\ket}[1]{| #1 \rangle}

\newcommand{\cre}[1]{\hat{#1}^{\dag}}
\newcommand{\ann}[1]{\hat{#1}}

\newcommand{\abs}[1]{\left|#1\right|}

\begin{document}

\title{Identifying the Bose glass phase}

\author{R. Pugatch}
\author{N. Bar-Gill}
\author{N. Katz}
\author{E. Rowen}
\author{N. Davidson}

\affiliation{Department of Physics of Complex Systems,\\
Weizmann Institute of Science, Rehovot 76100, Israel}

\begin{abstract}
Introducing disorder into the Bose-Hubbard model at integer
fillings leads to a Bose glass phase, along with the Mott
insulator and superfluid phases \cite{boseglassbig}. We suggest a
new order parameter: $\det(\rho)$, where $\rho$ is the one body
density matrix, which is nonzero only within the Mott-insulator
phase. Alongside the superfluid fraction $f_s$, it is then
possible to distinguish the three phases. The Bose glass phase is
the only phase which has vanishing $\det(\rho)$ and superfluid
fraction. The vanishing of $\det{\rho}$ in the Bose glass phase
occurs due to the partial fragmentation of the condensate into
localized fragments, each with zero superfluid response, which
implies the presence of unoccupied sites and hence the presence of
lines of zeros in the one body density matrix. In the superfluid
phase, $\det{\rho}$ vanish for another reason - due to the
macroscopic occupation of a single particle state. Finally, we
suggest the enhancement of the three body decay rate in the Bose
glass phase, as an experimental indicator for the presence of
localized fragments.
\end{abstract}

\maketitle

\section{Introduction}

 The onset of superfluidity in random media and its connection to the
existence of a condensate fraction has been an intriguing research
subject in the $^4$He community for many years
\cite{SuperfluidityInVycor,SuperfluidityBook}. In recent years,
ultra-cold atomic states in optical lattices have provided an
experimental means to answering these long-standing problems
\cite{blochMott}.

 The remarkable tunability and dynamical control of optical potentials 
offers the opportunity to emulate elaborate physical situations 
and create novel quantum many-body states of fundamental interest. 
In particular, the study of disordered systems comprises an important 
avenue of research. Specifically, the Anderson localization transition, 
perhaps the best known example for a quantum phase transition from 
an extended to a localized state, has never been unambiguously observed in an
experiment. Interacting bosons in a disordered potential can
present interesting and sometimes unexpected behavior, as we shall
discuss shortly. It is known \cite{SubirQPT,boseglassbig} that
interacting lattice bosons in the presence of a disordered on-site
potential exhibit a phase known as Bose glass, characterized as an
insulating phase with no gap in the excitation spectrum.

 The Bose-Hubbard Hamiltonian \cite{SubirQPT} with on-site disorder
is given by
\begin{equation}
H = - \frac{J}{2}\sum_{i} \cre{a}_{i} \ann{a}_{i+1}+h.c +
\frac{U}{2}\sum_{i} \cre{a_i} \cre{a_i} \ann{a_i}
\ann{a_i}+\sum_{i}W_i\cre{a_i} \ann{a_i}, \label{eq:BH}
\end{equation}
where $a_i$ is the annihilation operator for an atom at site $i$,
$J$ is the tunnelling strength, $U$ is the repulsive on site
interaction strength, and $W_i$ is the random on site potential,
drawn from a flat distribution centered around zero, of width
$\Delta$.

 When $W_i=0$ for all $i$, this Hamiltonian exhibits a
second order quantum phase transition from an energy gapped
Mott-insulator to a gapless superfluid state at integer filling
(number of atoms equal the number of sites), as the ratio $U/J$ is
changed \cite{SubirQPT}. The critical value for this transition
depends on the dimensionality of the underlying lattice.

 This model was extensively studied by a variety of methods, including
quantum monte-carlo in $1D$ and $2D$ \cite{QMC,BoseGlassMonte},
exact solutions in $1D$ for small lattices
\cite{Burnettboseglass}, renormalization group
\cite{RGBG_Example}, and recently in the context of cold atoms in
optical lattices using the mean field approximation
\cite{BoseGlassZoller}. One of the main conclusions of these
studies was that due to the disorder, there is a new
incompressible yet insulating phase, named the Bose glass phase.
It was conjectured by Fisher et. al. \cite{boseglassbig}, and
verified by \cite{Svistunov} and \cite{Rapsch} that any transition
from the Mott-insulator phase to the superfluid phase in the
presence of disorder, should follow through the Bose glass phase
(at integer filling).

 In this paper we suggest a new criterion for identifying the Bose glass phase, based
on the comparison between the superfluid fraction, and the
structure of the one body density function. We show that in the
presence of interactions, however small, a localized condensate
becomes unstable, and fragmentation occurs. In the thermodynamic
limit and a finite interaction strength, the number of localized
eigenfunctions scales with the size of the system, so the
occupation of each fragment (eigenfunction of the one body density
matrix) tends to zero. As we will explain below, one can use two
order parameters, the superfluid fraction $f_s$, and $\det{\rho}$,
where $\rho$ is the one body density matrix, to distinguish
between the three phases, since $f_s$ is non-zero only in the
superfluid phase and $\det(\rho)$ is non-zero only in the Mott
insulator phase (where $f_s=0$).

\section{Interplay between kinetic, potential and interaction energy}

In this section we will use a variational argument, to demonstrate
how the relative strength of the kinetic potential and interaction
energies, can stabilize a different ground state, either a Mott
insulator (strong interaction limit), Superfluid (both disorder
and interaction comparable to the kinetic energy), and Bose glass
phase (disorder or interaction or both are dominant).
Surprisingly, we will find regimes where by either increasing the
interaction strength $U$ or the disorder strength $\Delta$ we can
induce superfluidity in the system. The latter phenomenon deserves
a special name and is called disorder induced order (see e.g.
\cite{disorderinducedorderone,disorderinducedordertwo}). In what
follows, we devise a variational argument that compare three
typical limiting forms of the many body ground state, and show how
as a function of $U$ and $\Delta$ the nature of the ground state
changes in accord with the known phase diagram of the system. We
further point out how this result supports our suggested criteria.

 Consider first, a $1D$, non interacting, single particle disordered system with
the following Hamiltonian:
\begin{equation}\label{AG}
    H=-\frac{J}{2}(\sum_i \cre{a}_i\ann{a}_{i+1} + \ann{a}_i\cre{a}_{i+1}) +
    W_i\cre{a}_i\ann{a}_i.
\end{equation}
As before, $W_i$ is drawn randomly from a flat distribution of
width $\Delta$ centered around zero. All the eigenstates $\phi_i$
of this Hamiltonian are localized, and we denote the lowest energy
as $E_0$ and the corresponding eigenstate to be $\phi_0$, so
$H\phi_0 = E_0 \phi_0$.
\begin{figure}
  \includegraphics[width=9cm,height=5cm]{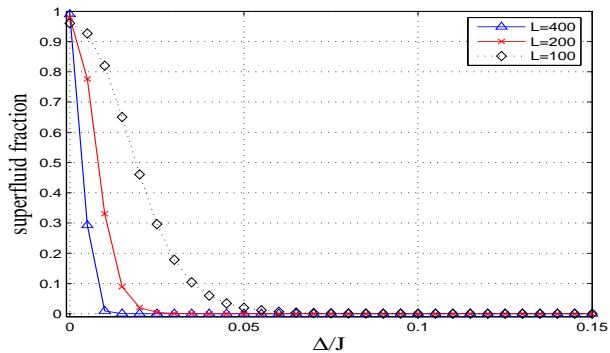}
  \caption{Superfluidity as a function of the disorder strength for
  different lattice sizes at zero interaction energy. Note that
  this superfluid fraction is the response of the many-body wave
  function to a phase twist in the boundary conditions, hence it
  is equal to $100\%$ for zero disorder and interaction strength,
  without contradicting the landau definition, which is a
  stability criterion \cite{Pines}.
}\label{sfVsdelta}
\end{figure}
Inserting into this system $N$ non interacting bosons, all of them
condense in the lowest energy state. Trivially,  the total many
body wave function is then given by:
$\psi_{AG}(x_1,...,x_N)=\prod_i\phi_0(x_i)$, where $AG$ stands for
Anderson glass. This state is an example of a completely localized
state, with zero superfluid response, but nevertheless, it is
macroscopically occupied. Some authors refer to this state, as an
Anderson glass \cite{AG1,AG2}.

 In Fig. $1$, we plot the superfluid fraction of this ground state, as a function
of the disorder strength for $L=100,200$ and $400$ sites. We
calculated the superfluid fraction $f_s$ exactly using the
following formula \cite{Burnettsuperfluidity}
\begin{equation}
    f_s=-\frac{1}{NJ}(-\frac{1}{2}\bra{\phi_0}\hat{\textbf{T}}\ket{\phi_0}-\sum_{i \neq 0}\frac{\bra{\phi_i}\hat{\textbf{J}}\ket{\phi_0}}{E_i-E_0}),
\end{equation}
where $\hat{\textbf{T}}$ is the kinetic energy operator and $\hat{\textbf{J}}$ is the current operator. In the following section we elaborate further on this definition of superfluidity.

 It is evident from Fig. $1$ that the critical point $\Delta_c$ where superfluidity disapear tends to $\Delta_c=0$ in the thermodynamic limit $L \rightarrow \infty$.
This is of course expected, since any disorder is strongly
localizing in $1D$ \cite{Localizationin_1d2d}. What happens if we
now turn on the interaction to a small, positive value? Assuming
the usual S-wave interaction, the interaction energy of the
localized state $\phi_0$ is $U_0N(N-1)/2$, where $U_0=g\int dx
\abs{\phi_0}^4$, and $g$ is the effective $1D$ interaction
coefficient \cite{Shlep}. This energy is much larger than the
energy of a fragmented state of the form $\psi_{BG}=\prod_i
{\phi_i(x_i)}^{n_i}$ where the integers $n_i$ are zero above some
$k\leq N$, and $\sum_i n_i = N$ (as we will see shortly, the case
$k=N$ corresponds to the Mott insulator phase), and $\phi_i(x)$
are the eigenstates of the free particle Hamiltonian in Eq.
\ref{AG}. The interaction energy of this state is approximately
given by $\sum_i U_i n_i(n_i-1)/2+\sum_{i\neq j} U_{ij} n_i n_j$,
where $U_i=g\int dx \abs{\phi_i}^4$ is the inter state interaction
energy, and $U_{ij}=g\int dx \abs{\phi_i}^2\abs{\phi_j}^2$ is the
intra state interaction energy\cite{remark}. It is not difficult
to find a set of $n_i$'s which gives a smaller energy than a
macroscopically occupied condensate wave function of the form
$\psi_{AG}$. To do so, we start by putting the first particle in
the lowest energy state $\phi_0$ we then add the next particle,
searching among all states for the state that will minimize the
total energy (including interaction). We repeat this process until
we fill the lattice with $N=L$ particles, to obtain a many-body
state that has the following form:
\begin{equation}
\psi_{BG}(x_1,...,x_N) = \prod_i \phi_i(x_i)^{\bar{n}_i}. \label{BGansatz}
\end{equation}
Hence, by this method we obtain a specific set of occupation
numbers $\bar{n}_i$'s, some of which are zero. We calculated the
$\phi_i$'s for a lattice with $500$ sites, which is effectively
thermodynamic in the regime of $\Delta$'s and $U$'s that we
employ.

 Next, we construct two more many body wave functions: $\psi_{SF}(x_1,...,x_N)=L^{-N/2}$ which corresponds to the
superfluid state in the absence of disorder and interactions,
since the condensate wave function then is $1/\sqrt{L}$, and all
$N$ particles occupy this extended state. The second wave function
is $\psi_{MI}(x_1,...,x_N)=\prod_i e(x-x_i)$, where $e(x-x_i)$ is
$1$ at site $i$ and zero otherwise. This state approaches the Mott
insulator (MI) ground state for large $U's$ and no disorder.

 In Fig. $2$ we plot the total energy of each of these four many boy wave functions:
$\psi_{AG},\psi_{BG},\psi_{SF}$ and $\psi_{MI}$ as a function of
the interaction strength $U$, for $\Delta/J=2$. The total energy
of each of the following states is given by: $E_{BG}=\sum_i n_i
E_i+U_i n_i(n_i-1)/2+\sum_{i \neq j} U_{ij} n_i n_j$, where $E_i$
is the eigenstate of the free single particle Hamiltonian $H_0$,
$E_{SF}=U_{SF} N (N-1)/2-2(N-1)J +Tr(H_0)$, is the single body
extended state, $L$ is the length of the lattice, $U_{SF}=g/L$,
and $H_0$ is the free particle Hamiltonian, $E_{MI}=Tr(H_0)-2NJ$,
\\and $E_{AG}=NE_0+g\int dx\abs{\phi_0}^4 N(N-1)/2$.
\begin{figure}
  \includegraphics[width=8cm,height=7cm]{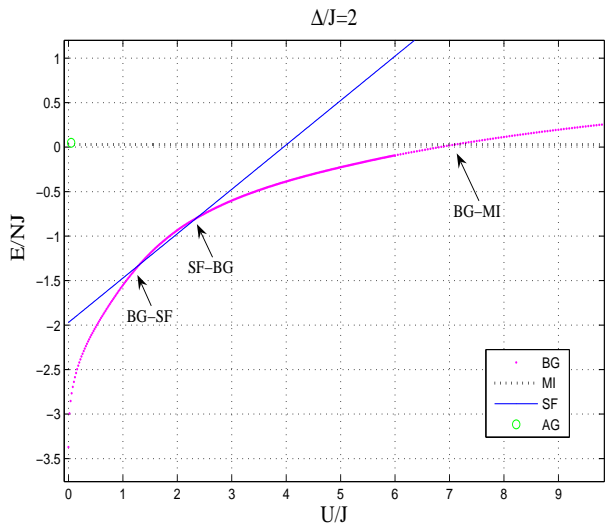}
  \caption{Energies per particle (in units of $J$), of the four variational many body states: $\psi_{SF}$, $\psi_{BG}$, $\psi_{MI}$ and $\psi_{AG}$, as a function of the interaction strength $U/J$, for
$\Delta/J=2$. The $\psi_{AG}$ wave function has an energy which is
larger then all the other states, for all non-zero $\Delta$'s.
From the graph it follows that the minimal state is ordered
$\psi_{BG}$,$\psi_{SF}$,$\psi_{BG}$,$\psi_{MI}$, as expected. This
shows that interactions tend to delocalize at small $\Delta$'s,
since they penalize localization to specific sites on the one
hand, and the gain in kinetic energy of being delocalized is high,
on the other hand. For larger interaction strength, this is no
longer true, and a superfluid is formed. At even larger $U$'s the
Bose glass phase is re-entered. Finally, when the interactions are
even stronger, the particles will minimize the energy just
localizing at the lattice sites, disregarding the disordered
potential.}\label{VariationalArgument}
\end{figure}

 We find that there are $4$ regimes: for $\Delta/J<1.3$ The minimal energy state corresponds to $\psi_{BG}$.
For $1.3<\Delta/J<2.35$ we find that the minimal energy state is
$\psi_{SF}$, which corresponds to the superfluid phase. As
expected, in this regime, the condensate fraction is large, and
the condensate wave function is extended. For $2.35<\Delta/J<7.1$
the $\psi_{BG}$ phase is again minimal. Finally, for
$\Delta/J>7.1$ the state $\psi_{MI}$ has minimal energy. Comparing
these approximate results to the known phase diagram we conclude
that this simple variational argument already exhibit the rich
interplay between the kinetic energy, the disordered potential
energy and the interaction energy by displaying the right ordering
of the phases. Starting from the Bose glass phase and increasing
the interactions, the superfluid phase can become favorable. For
larger interaction strength, the Bose glass phase once again
stabilizes, and for even larger interactions, finally, the Mott
insulator phase wins. This correctly depicts the true behavior of
the ground state \cite{Rapsch}. Also, we note that the localized
condensate is unstable for any non-zero interaction strength,
hence, the Anderson glass in practice, is a phase which strictly
speaking can only exist on the $U=0$ line. Indeed, when
interactions are present, the energy cost of concentrating all the
particles inside the localization length, grows without bounds in
the thermodynamic limit.

We note that the one body density function $\rho$ derived from
$\psi_{SF}$ has one macroscopic eigenfunction, but since
$Tr(\rho)=N$, it follows that some of the eigenstates are zero
(since the size of $\rho$ is $N$ by $N$), this is also true when
the macroscopic occupation of the extended state is not total due
to quantum depletion. The structure of $\rho$ for the $\psi_{BG}$
state is different, since instead of having long range order, the
density matrix is composed of small blocks, separated by lines of
zeros, indicating that there are lattice sites which are not
occupied, due to their large potential energy. In both cases, the
determinant of the density matrix is zero. Deep in the Mott
insulator phase, the determinant is one, since the one body
density function is the identity matrix. Near the transition,
still in the Mott insulator phase, the ground state wave function
is approximately $\ket{1,1,...,1}+(J/U)\sum_i \epsilon_i
\cre{a}_{i+1} \ann{a}_{i}\ket{1,1,...,1}$ which also has
$\det(\rho)>0$, $\epsilon_i\leq 1$ and are equal only when
$\Delta=0$. We therefore assert that $\det(\rho)$ is non zero in
all the $MI$ phase. This observation is just another way to
express the fact that only in the Mott insulator phase, a gap is
formed. When the number of occupied eigenstates (fragments) of
$\rho$ equals the number of particles, this gap will be
proportional to $U$, which is just the cost of moving a particle
to an occupied fragment.


\section{Definition of condensate and superfluid fractions}

 Following \cite{Legget}, we define the condensate fraction as the
properly normalized largest eigenvalue of the one body density
matrix $\rho(x,x')$. For example, when a single particle wave
function $\chi(x)$ exists such that
$\rho(x,x')=N\chi(x')^{*}\chi(x)$, then the condensate fraction is
$N$ - the number of particles in the system. Note that this
definition does not imply any specific characterization of the
condensate wave function $\chi(x)$, which can be extended or
localized.

 Superfluidity is manifest whenever there is a
difference in the momentum density response to a transverse vs. a
longitudinal imposed velocity field
\cite{Leggetsuperfluidity,BaymSuperfluidity,HuangSFArticle}. Here,
we employ the phase twist definition
\cite{StifnessDefSupefluidity,Burnettsuperfluidity}, where we
impose a phase twist $\theta$ along one of the dimensions of the
system, corresponding to the application of a transverse velocity
field \cite{BaymSuperfluidity}. We equate the superfluid velocity
$v_s$ with $\frac{\hbar}{m}\frac{\theta}{L}$, where $m$ is the
effective mass of the atom, and $L$ is the length of the system
\cite{BaymSuperfluidity}. If the phase twist is small, then the
difference in the ground state energy between the twisted
($E_{\theta}$) and untwisted ($E_0$) systems can be attributed to
the kinetic energy of the superfluid. The superfluid fraction
$f_s$ is therefore given by $(1/2)mv_s^2f_sN=E_{\theta}-E_0$, with
$N$ the total number of atoms. For a normal fluid, this difference
is zero, indicating an identical response to longitudinal and
transverse imposed velocity fields.

\section{Studying the exact solution}

 Using an exact solution for $8$ particles in $8$ sites,
we show that with this criterion we can roughly obtain (up to our
finite size resolution) the correct phase diagram of the system.

\subsection{A lattice with one defect}

 The effect of a random potential on an interacting bosonic system
can be identified by measuring the superfluid fraction. In
practice, for a finite size system with a longitudinal extension
$L_z$, residual superfluidity is expected to persist as long as
$\xi_{loc} > L_z$.
 Remembering that in $1D$ any disorder is localizing, we first consider
the simplest kind of disorder, namely, a single site defect, by
adding a negative external potential at one particular site, with
strength $\Delta$. In Fig. $3$, we show the result of an exact
diagonalization of Eq. (\ref{eq:BH}) we performed, for a $1D$
system with $8$ atoms and $8$ sites, and a varying defect strength
$\Delta$ (in units of $J$). From the figure it is apparent that
starting from the superfluid regime ($U/J=1$) and increasing
$\Delta$, the superfluid response decreases rapidly to zero, while
the condensate fraction increases to $100\%$.
\begin{figure}[h]
    \centering
        \includegraphics[width=7cm,height=4cm]{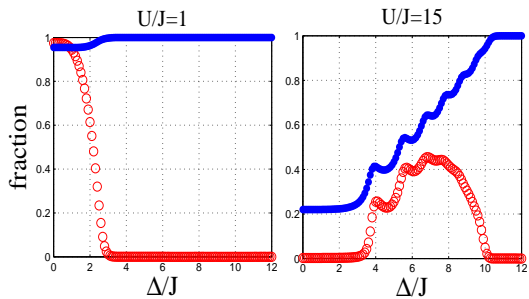}
  \caption{Exact calculation of the largest eigenstate of $\rho$ (filled circles) and
the superfluid fraction (hollow circles) for a one dimensional
lattice with $8$ atoms and $8$ lattice sites, and a single site
defect potential $-\Delta$. (a) In the weakly interacting case
($U/J=1$), increasing $\Delta/J$ causes all the atoms to enter the
defect, so the maximal eigenvalue of the one body density is
$100\%$ of the particles, while the superfluid fraction decrease
rapidly to zero, as the condensate wave function becomes
localized. (b) Interactions are stronger ($U/J=15$), so at first,
lowering the defect causes delocalization of the atoms, hence,
superfluidity becomes possible. Only at higher values of
$\Delta/J$ it is energetically favorable to become localized at
the defect, and superfluidity disappears.}
\label{fig:ToyModelGraph}
\end{figure}
As we lower the defect, atoms localize there, without the ability
to participate in the superfluid flow. When all the atoms are
captured by the defect, they all occupy the same single particle
wave function which is localized at the defect site. Starting, on
the other hand, from the Mott insulator phase ($U/J=15$), and no
disorder ($\Delta=0$),
essentially all the atoms are pinned to the lattice sites. 
Increasing $\Delta$ causes an onset of the superfluid fraction,
since in the presence of large interactions atoms can lower their
energy by being in an extended state, hence making superfluidity
possible. This will be the basis for understanding the more
elaborate statement that one can onset superfluidity by increasing
the disorder strength - disorder induced order. Only for larger
$\Delta$, the potential energy can compensate for the gain in
interaction and kinetic energies by localizing again at the defect
site, hence losing superfluidity as indicated in the figure for
$\Delta/J>10$. We confirm that the condensate wave function is
indeed extended (comparable non-zero values throughout the
lattice) whenever superfluidity is present
\cite{GriffinTalbot,remark}.

 In this section we refereed to the condensate fraction, since for a single impurity which is properly scaled
it can remain finite even at the thermodynamic limit, whereas (as
will be shown in the next section) for a disordered potential the
condensate fraction is non zero only for a small (mesoscopic)
system, and tends to zero in the thermodynamic limit.

\subsection{Exact solution with a random on-site potential}

 Next, we return to the disordered potential, which
acts on all $8$ sites. We exactly diagonalize Eq. (\ref{eq:BH})
and calculate the determinant of the one body density matrix and
the superfluid fraction, for different values of $U/J$ and
$\Delta/J$, where now, again, $\Delta$ is the width of the flat
distribution from which we draw the random values of the on-site
potential. The results are shown in Fig. $5$.
 Looking at the superfluid fraction as a function of $\Delta$ for $U=0$, we see that there is a
rapid decrease of superfluidity towards zero due to the strong
localization (the superfluidity starts at a non-zero value due to
the finite size of our lattice). Looking at the dependence on $U$
for $\Delta=0$, on the other hand, we observe the usual superfluid
to Mott-insulator transition, which is slightly smeared due to
finite size. It is also evident from Fig. $5$ that $\det{\rho}$ is
non zero only at the Mott insulator phase.

 There are two somewhat surprising features apparent in Fig. $5$. First, starting from small $\Delta/J$, and
increasing the interactions, we observe the onset of
superfluidity, indicating that in this regime, interactions tend
to delocalize. Second, starting from the Mott-insulator phase (e.g
at $U/J=8$), and increasing the disorder strength, induces
superfluidity, a phenomenon called \emph{disorder induced order}
\cite{disorderinducedorderone}.

 By inspecting the eigenvalues of $\rho$ we verified that in the regimes where
the superfluid fraction $f_s$ is zero, the condensate is
fragmented, i.e. the number of non-zero eigenvalues of $\rho$ is
larger than one. While both the delocalizing role of $U$ and
disorder induced order are already present in the single defect
model discussed above (see e.g. Fig. $3$), fragmentation can not
occur for single site ``disorder'' since there is always an
overlap between any extended state and a localized state.

 Comparing these results to the known phase diagram \cite{Rapsch},
we find that, whenever the Bose glass phase is present, at least
one but less than $I$ eigenvalues of $\rho$ ($I$ being the total
number of sites), are non-zero while the superfluid fraction is
zero. In the Mott insulator, we observe complete fragmentation,
and all the eigenvalues are non-zero (with $f_s=0$). Finally, in
the superfluid phase, some of the eigenvalues are zero, while the
large eigenvalue is large (of the order of $1$).

 Further insight is gained by looking at specific eigenfunctions of $\rho$.
To avoid the limitations of finite size, we inspect the Bose glass
and the Mott insulator phases in a regime in the $\Delta/J - U/J$
plane, where the localization length is $\lesssim 2a$ which is
less than the length of the system $I=8a$, where $a$ is the
lattice spacing. In Fig. $4$ we plot the eigenfunctions of $\rho$
for the Mott insulator phase (MI),  The superfluid phase, and the
Bose glass phase. We note that In the superfluid phase
($U/J=\Delta/J=1$), the largest eigenvalue of $\rho$ corresponds
to $95\%$ of the particles, and the accompanying eigenstate is
extended (comparable non zero values anywhere on the lattice). The
last eigenvalue of $\rho$ are zero i.e. they are unoccupied. In
the Mott insulator state ($U/J=15,\Delta/J=1$), we see that all
the eigenvalues are non-zero (the largest eigenvalue equals $22\%$
of the total number of particles, the second largest is $17\%$,
and the smallest is $7\%$), indicating complete fragmentation.
Also note that the eigenfunctions are all extended. Finally, In
the Bose glass regime, all the eigenfunctions are localized and
some of the eigenvalues of $\rho$ are zero, indicating partial
fragmentation. In particular, due to the finite size, when the
disorder is strong and the interactions are weak
($\Delta/J=15,U/J=1$), we find that the largest eigenvalue is
$97\%$ and the eigenfunction is localized, with $\xi \sim 2a$.
When both the interactions and the disorder are strong
($U/J=15,\Delta/J=60$), partial fragmentation occurs. The largest
eigenvalues of $\rho$ is then $58\%$, the second, third and fourth
largest are respectively: $17\%$,$13\%$ and $12\%$ and the rest
are zero. The results presented here are of a singe realization of
the disorder, which we choose as representative from a larger set
of disorder realizations, all with similar quantitative results.

\begin{figure}
    \centering
        \includegraphics[width=6cm]{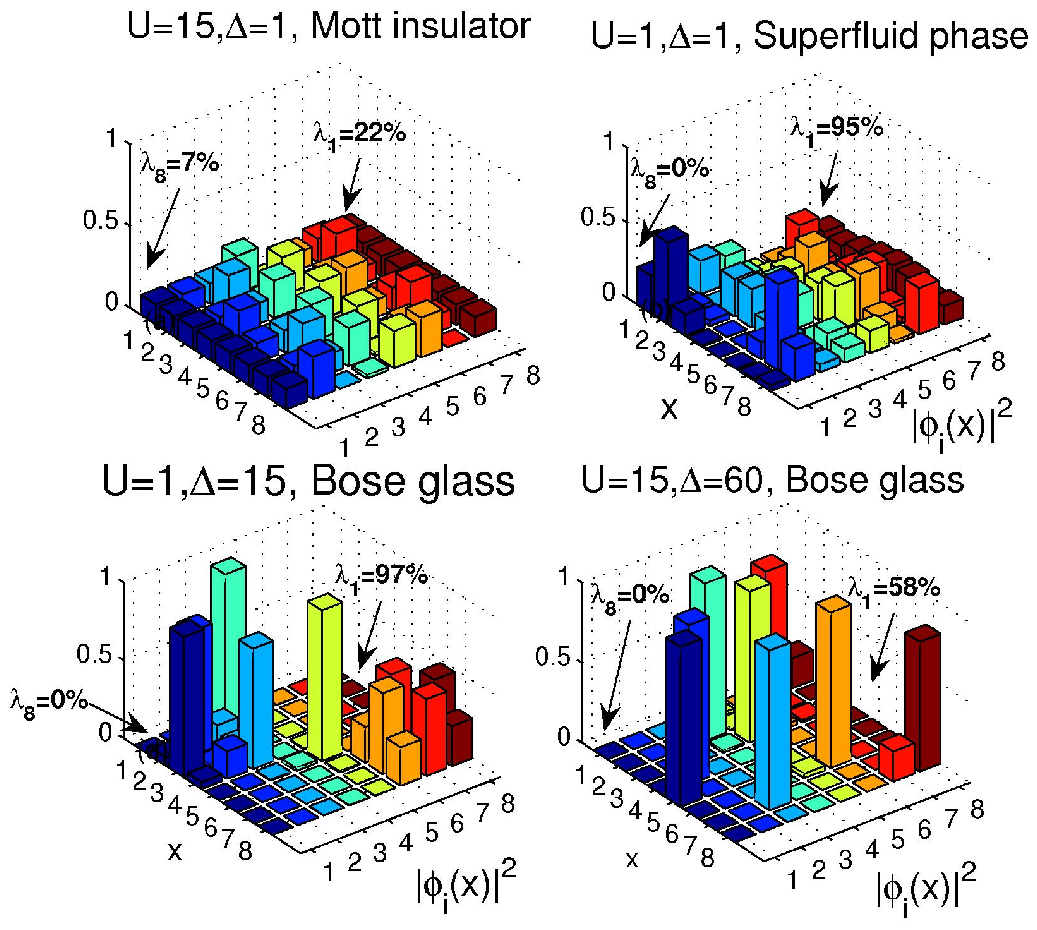}
  \caption{Absolute value squared of the eigenfunctions $\eta_i(x)$ of the one body density
matrix $\rho_{ij}=\bra{g.s.}\cre{a}_i \ann{a}_j \ket{g.s.}$, as a
function of position (in units of lattice spacing). $\lambda_i$ is
the corresponding eigenstate of $\eta_i$, indicated only for the
largest ($i=1$) and smallest ($i==8$) eigenvalues.}
\label{fig:example}
\end{figure}

\section{A new order parameter for the Mott insulator phase}

 A single, base invariant, order parameter that is
non-zero only in the Mott insulator phase is $\det(\rho)$. Since
$Tr(\rho)=N$, it follows that only when there is exactly one
particle in each eigenfunction, all the eigenvalues of $\rho$ are
equal to $1$ then $\det(\rho)=1$, which happens in the limit $U
\rightarrow \infty$. In the superfluid phase and in the Bose glass
phase, $\det(\rho)=0$. We therefore propose that the combination
of the superfluid fraction and $\det(\rho)$ can distinguish the
three different phases.

Inspecting $\det(\rho)$ as a function of $U/J$ for $\Delta=0$, we
find that the superfluid to Mott insulator transition occurs at
$~4.6$ in accordance with \cite{Burnettboseglass}. We also
verified for large $U$ $\det(\rho)$ tends to unity (for example at
$U=1000$, $\det(\rho)>0.9999$).

\begin{figure} [b]
    \centering
        \includegraphics[width=7cm]{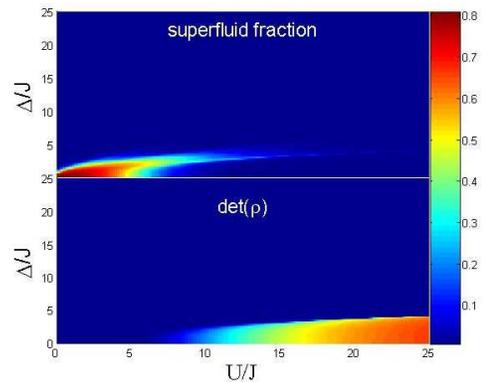}
  \caption{Superfluid fraction, and $\det(\rho)$ as a function of $U/J$ and $\Delta/J$.
  The Bose glass phase is characterized by zero superfluidity and $\det(\rho)=0$, The Mott insulator phase is the phase
  with non zero $\det{\rho}$ and zero superfluidity, and the superfluid phase naturally has non zero superfluidity and $\det{\rho}=0$}
\label{fig:newOP}
\end{figure}


 Thus, the Bose glass phase can be thought of as a phase composed of
localized eigenstates of $\rho$, without phase relations between
them. The total many body wave function is therefore, to leading
order, $\psi(x_1,...,x_n) \approx \prod_i \phi_i(x_i)^{n_i}$,
where $\sum n_i = N$ and some of the $n_i$'s are larger than $1$.

  In the presence of interactions, a single localized condensate at the lowest
point in the potential is unstable and will tend to be fragmented
to microscopic fragments (for large $\Delta/J$) or to delocalize
(for small $\Delta/J$). This is because the interaction energy is
proportional to the density squared, and thus whenever
interactions are  present, the ground state wave function will
take a form that has the lowest possible density. A single
extended condensate is formed (hence invoking superfluidity), when
$U/J$ is less than the homogeneous Mott insulator-superfluid
transition (so the energy penalty for occupying more than one atom
per site, is small compared to the kinetic energy gain) and also
$\Delta/J$ is small (so the penalty for occupying higher peaks in
the disordered potential is also small). Alternatively,
fragmenting to a set of non-overlapping localized condensates,
each centered at a different localization center, with negligible
energy differences (in the thermodynamic limit), is favorable when
$\Delta/J$ is large.

 Hence we expect that in the Bose glass phase, the number of
fragmented condensates $N_{loc}$ will generally be larger than
one. Moreover, $N_{loc}$ should scale like $L^{d}$ where $d$ is
the dimension of the underlying lattice. In the Bose glass phase
$N_{loc}<N$, since if $N_{loc}=N$, i.e. the number of fragments
equals the number of sites, then an energy gap will form, whereas
the Bose glass phase is not gapped. When the number of
localization centers is a finite fraction of the total number of
sites - a mesoscopic system, the number of atoms per localized
condensate will also be a finite fraction from the total, so the
condensate fraction will be non-zero. Moreover, in a finite
system, we expect a regime of parameters (arbitrary disorder and
weak interactions), in which the finite gap in the excitation
spectrum (that exists due to the finite size) will protect a
single localized condensate from fragmenting. This regime was
called by several authors an Anderson glass \cite{AG1,AG2}. We
stress that in the thermodynamic limit, the Anderson glass only
exist along the line $U=0$ (zero interactions).


 Due to the limited number of lattice sites and atom number in our exact calculation, in
Fig. $5$, and contrary to the variational calculation presented in
section two, we could not resolve the Bose glass phase intervening
between the Mott insulator and the superfluid phases
\cite{boseglassbig,Svistunov,Rapsch}. Keeping this limitation in
mind, we were able to recover the correct structure of the phase
diagram, as compared to previous work done with a different tool
\cite{Rapsch}.
 Our results can also be scaled to larger numbers. To demonstrate this, we repeated the calculations for
$N=6$ atoms and sites, and obtained a qualitatively similar phase
diagram to that with $N=8$. Moreover, using finite size scaling in
$1D$ in the limit of $\Delta=0$, we recover the Mott insulator
transition at $U_c = 4.85J$, in good agreement with the value
$4.65J$ presented in \cite{Burnettboseglass}. We also checked the
half filling case for $6$ atoms and $12$ lattice sites, where
since $N<I$, $\det(\rho)=0$ indicating the absence of the MI phase
as expected. Finally, we note that the fact that in the presence
of disorder a Bose glass will always intervene between the Mott
insulator and superfluid phases, seems more evident using our
criterion, since due to the breakdown of translational invariance,
the only way from the superfluid phase, where a macroscopic
condensate exists (no fragmentation), to the Mott insulator state,
where all the eigenfunction of $\rho$ are occupied (complete
fragmentation), goes through gradual partial fragmentation.

\section{Using the three body loss to identify the superfluid to Bose glass transition}

 In this section, we propose a new experimental indicator for the presence of
localized states within the Bose glass phase, \emph{when
interactions are small}. To do so, we propose that for a
mesoscopic system such as the systems described in
\cite{TonksGasBloch,BG_Ingucio}, it is possible to use the three
body decay rate \cite{ThreeBodyLoss,ThreeBody}, to identify the
presence of localized states. We consider a collection of $1D$
trapped BEC's created by loading a $3D$ BEC into a deep
two-dimensional optical lattice, and introduce a controlled
disorder of strength $\Delta$ relative to $J$ (for a more
accessible experimental scheme see
\cite{MeAndBarGill,BG_Ingucio}). The $1D$ three body loss rate
coefficient is $\Gamma^{1D}=\Gamma^{3D}/12\pi^2\sigma_{r}^{4}$,
where $\Gamma^{3D}=5.810^{-42} m^6/sec$ is the corresponding $3D$
three body loss rate coefficient for $^{87}Rb$ \cite{ThreeBody},
$\sigma_{r}=85 nm$ is the assumed radial confinement, and $L=150
\mu m$ is the condensate length \cite{TonksGasBloch}. Hence, we
obtain that $\Gamma^{1D}=9.4 10^{-16} m^2/sec$. In Fig. $6$, we
plot the total three body decay rate $\Gamma_{tot}$ as a function
of the disorder strength $\Delta$ relative to $J$, for different
values of $U/J$, according to
\begin{equation}
    \Gamma_{tot}=\sum_i \Gamma^{1D}
    \bra{g.s.}\cre{a}_i\cre{a}_i\cre{a}_i\ann{a}_i\ann{a}_i\ann{a}_i\ket{g.s.}/L^3.
\label{three body loss}
\end{equation}
\begin{figure} [t]
    \centering
        \includegraphics[width=9cm,height=5cm]{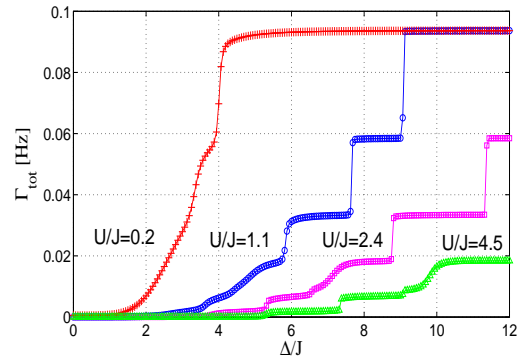}
\caption{Exact calculation of the total three body decay rate as a
function of $\Delta/J$ for different values of the $U/J$
(presented in units of $(U/J)_c$), for $N=I=8$ particles and
sites. Starting at the superfluid phase and increasing $\Delta$,
localization increase the density and hence the three body decay
rate sharply increases. When $U/J$ is large, the decay rate is
still notably larger in the Bose glass phase compared to the
Mott-insulator and superfluid phases, but it is less than the
decay rate for smaller values of $U$, due to larger fragmentation
that decreases the density. When all the particles are localized
at the lowest defect, the total decay rate reaches its maximal,
asymptotic value, which here is small due to small size of the
system (compare with Fig. $7$.}
    \label{fig:three body loss}
\end{figure}
 Looking at Fig. $7$ together with Fig. $5$, we observe the following features. When the system is in the Mott
insulator phase, the total three body decay rate $\Gamma_{tot}$ is
negligible, since the amplitude for three atoms to occupy the same
site is effectively zero. In the superfluid phase, the bosons are
now delocalized across the lattice. As a result the amplitude to
occupy a single site with three or more atoms increases to a small
value, and $\Gamma_{tot}$ also increases to a small value. The
most dramatic effect is seen when we enter the Bose glass phase
and the interactions are weak. In this regime, due to the the
existence of localized condensates with increased density, we
observe a notable increase in $\Gamma_{tot}$. When the number of
fragmented condensates increases, we see a corresponding decrease
in the loss, as expected. The step-like behavior of $\Gamma_{tot}$
is due to the fact that the capture of another particle is a
notable event, since the total number of particles is small.
\begin{figure}
    \centering
        \includegraphics[width=9cm,height=5cm]{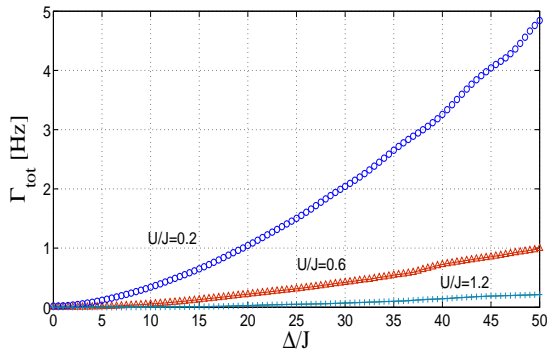}
\caption{Mean field calculation of the total three body decay rate
as a function of $\Delta/J$ for different values of the $U/J$
(presented in the corresponding $(U/J)_c$ units), for $101$ atoms
and sites. Similar to the exact result of Fig. $6$, we observe
that when the interactions are weak, the increase in the three
body loss rate as a function of the disorder strength $\Delta/J$
is pronounced. The decay rate decreases, when the interactions
increase, since the presence of interaction cause the density to
decrease due to fragmentation.}
    \label{fig:three body loss_MF}
\end{figure}

 In Fig. $7$ we repeat the analysis, using
a novel mean field simulation based on the Gutzwiller
approximation \cite{stoofopticallattices,Gutzwilleransatz}, that
we developed, which is capable of working at a constant filling
factor for arbitrary external potentials. With this simulation, we
also calculated an upper bound for the superfluid density, and
recovered qualitatively the upper part of Fig. $5$. In Fig. $7$,
we calculated the total decay rate from a $1D$ lattice with $101$
atoms and sites. We calculated the density profile in the presence
of disorder, and the approximate ground state of the system, which
we insert to Eq. \ref{three body loss} to obtain the loss rate. We
note that although the simple Gutzwiller ansatz is unable, by
construction, to recreate fragmentation, the density, profile is
expected to be accurate for our purposes. The results, shown in
Fig. $7$, are consistent with the exact results, up to a trivial
scaling of $U/J$ and $\Delta/J$ due to the change in the critical
point in the mean field model. They support our claim that three
body loss is a good measure for localization in a mesoscopic
system, in the regime where interactions are weak.

 Finally, we note that localized condensates may also be identified by careful examination of their
momentum distribution observed in time-of-flight images
\cite{Burnettboseglass}. This can be considered as a complementary
measure, to the three body loss, since as interactions increase,
fragmentation becomes larger hence the loss rate decreases, while
the interference is washed out.
\\
\section{conclusion and outlook}

 In this paper we propose a new criterion for identifying the Bose glass phase, as a phase
composed of localized fragmented eigenfunctions of $\rho$ (with
some of the localized eigenfunctions remaining unoccupied), and
zero superfluidity. Starting from the Bose glass phase and
increasing the interaction strength can either induce a transition
to the superfluid phase, where there is one macroscopically
occupied \emph{extended} condensate, or to induce further
fragmentation. When the number of fragments equals the number of
sites, we enter the Mott insulator phase, which also has zero
superfluidity, but with a non-zero energy gap. A summary of these
observation is encapsulated by the second order parameter
$\det{\rho}$ that we propose, alongside the superfluid fraction
$f_s$, which is non zero only in the Mott insulator phase. We
showed that $\det(\rho)$ is non zero in the Mott insulator side,
both near and far from the transition to the Bose glass phase. We
also argued that we can capture some of the features of the
transition, in particular, that the Bose glass phase is reentered
as $U$ is increased, before entering the Mott insulator phase, by
using a variational ansatz, which also obeys our criterion. We
find for an exact solution for a small system with $8$ particles
and sites, that using our suggested criterion we can obtain the
correct phase diagram (up to our limited finite size resolution),
and that fragmentation mechanism occurs for large $U$'s.

 Finally, we discuss the possibility to observe
localization in a realistic experimental setup in the regime of
small interaction strength, using the increase in the three body
decay rate. Several intriguing effects resulting from the
interplay between disorder and interactions such as disorder
induced order, arise already in a single defect scheme, which may
be more accessible experimentally.

 This work was supported in part by the Israel Science Foundation, and by the Minerva
Foundation.
\bibliography{bib3}

\begin{thebibliography}{34}
\bibitem[{rem()}]{remark}
\bibinfo{note}{This approximation works best for weak interactions, since in
  this regime the noninteracting basis is still appropriate.}
\end{thebibliography}
\end{document}